\documentclass{iopjournal}

\makeatletter

\renewcommand\appendix{%
  \par
  \setcounter{section}{0}%
  \setcounter{subsection}{0}%
  \setcounter{subsubsection}{0}%
  \setcounter{equation}{0}%
  \setcounter{figure}{0}%
  \setcounter{table}{0}%
  \gdef\thesection{Appendix \Alph{section}}%
  \gdef\theequation{\Alph{section}.\arabic{equation}}%
  \gdef\thefigure{\Alph{section}\arabic{figure}}%
  \gdef\thetable{\Alph{section}\arabic{table}}%
}

\makeatother

\usepackage{lmodern}
\usepackage[T1]{fontenc}
\usepackage{caption}
\usepackage{subcaption}
\usepackage{float}
\usepackage[parfill]{parskip}
\usepackage{amsmath,amsfonts,amssymb}
\usepackage[utf8]{inputenc}
\usepackage{booktabs}
\DeclareUnicodeCharacter{2208}{\in}
\usepackage{makecell}

\begin{document}


\title{Molecular electrostatic potentials from machine learning models for dipole and quadrupole predictions}

\author{Kadri Muuga$^1$\orcid{0000-0000-0000-0000}, Lisanne Knijff$^{1,2}$ and Chao Zhang$^{1,3*}$\orcid{0000-0002-7167-0840}}

\affil{$^1$Department of Chemistry-Ångström Laboratory, Uppsala University, Lägerhyddsvägen 1, P.O. Box 538, 75121 Uppsala, Sweden }

\affil{$^2$Current address: PDC Center for High Performance Computing, KTH Royal Institute of Technology, SE-10044, Stockholm, Sweden}

\affil{$^3$Wallenberg Initiative Materials Science for Sustainability, Uppsala University, 75121 Uppsala, Sweden}

\affil{$^*$Author to whom any correspondence should be addressed.}

\email{chao.zhang@kemi.uu.se}

\keywords{Dipole moment, quadrupole moment, electrostatic potential, machine learning, AI}

\begin{abstract}
The molecular electrostatic potential (MEP) is a key quantity for describing and predicting intermolecular and ion–molecule interactions. Here, we assess the ability of machine-learning (ML) models to infer the MEP, based on the equivariant graph-convolutional neural network architecture PiNet2 and trained on dipole and quadrupole moments. For the established QM9 dataset, we find that including the quadrupole contribution in the ML models substantially improves their ability to recover the MEP compared to dipole-only models. This trend is confirmed on the SPICE dataset, which spans a much broader region of organic chemical space. Together, this study underscores the central role of the quadrupole moment as a fitting target for ML models aiming at rapid access to the MEP.
\end{abstract}

\section{Introduction}

Molecular electrostatic potentials (MEPs) provide a compact way of describing how a molecule interacts electrostatically with its environment. MEPs have long been used to rationalize sites of electrophilic and nucleophilic attack, hydrogen bonding patterns, molecular recognition, (electro)catalytic reactivity and solvent design~\cite{10.1515/pac-2025-0555, WCMS:WCMS19, 10.1002/aenm.202300259, 2023.Zhang4uc}. Nevertheless, one rarely works directly with continuous MEPs in practical molecular  simulations. Instead, the potential is represented in terms of a set of atomic charges that are tuned to reproduce the \emph{ab initio} electrostatic potential (ESP) around the molecule as faithfully as possible~\cite{10.1039/c0cp00111b}.

This idea gave rise to the “ESP-charge” family of methods, in which atom-centered monopoles are obtained by fitting the quantum-chemical MEP on a grid of points surrounding the molecule. Well-known examples include the Merz–Kollman scheme~\cite{besler_merz_semiempirical}, CHELP/CHELPG~\cite{JCC:JCC540110311}, and the restrained electrostatic potential (RESP) model~\cite{bayly_well-behaved_1993}. These methods differ mainly in how the grid points are chosen, how rotational invariance and numerical stability are enforced, and how regularization or restraints are applied to avoid unphysical charges for highly buried atoms.

One such regularization is to supplement the ESP fit with constraints that enforce the correct multipole moments~\cite{sigfidsson_ulf_esp_chg_method_comparison}. This improves the description of the far-field electrostatics and helps stabilize the fit for flexible or highly polar molecules. A different but related avenue is exemplified by charge model 5 (CM5)~\cite{Marenich:2012et}, which starts from Hirshfeld population analysis and then maps the resulting atomic charges through element- and environment-specific corrections that are parametrized to reproduce gas-phase dipole moments across a large and diverse training set.

In parallel with these developments, atomistic machine-learning (ML) approaches have emerged as a powerful way to construct environment-dependent charge models. A prominent example is the work by Sifain and co-workers~\cite{Sifain:2018fr}, who trained a deep neural network to assign atomic charges such that the resulting dipole matches high-level electronic-structure references. Their affordable charge assignment (ACA) model produces charges that closely resemble CM5 in magnitude and trends, which was confirmed later by other groups~\cite{Shao:2020kra}. Given the well-known expression between the MEP and the multipole moments~\cite{jackson_classical_1999}:
\[V(\textbf{r}) = \frac{1}{4\pi \epsilon_0} \left( \frac{q_\textrm{tot}}{|\mathbf{r}|} + \sum_\alpha \frac{\mu_{\alpha} \cdot r_{\alpha}}{|\mathbf{r}|^3} +\sum_\alpha\sum_\beta\frac{Q_{\alpha \beta} \cdot r_{\alpha}r_{\beta} 
}{2|\mathbf{r}|^5} + ...\right) \label{multipole_exp} \tag{1}  \]

where $\alpha$ and $\beta$ refer to the x, y and z-coordinates. The following two questions naturally arise: 1) how good would the ability of ML models trained for dipole predictions be to infer MEP? 2) would the inclusion of quadrupole contribution in the ML models improve this ability? 

In this work, we have expanded the PiNet-dipole model (AC dipole model hereafter) developed previously~\cite{Shao:2020kra, knijff_machine_2021} to include the contribution from quadrupole moment. This was achieved by exploring the newly updated equivariant graph-convolutional neural network architecture PiNet2~\cite{pinet2} and by labelling the QM9 dataset for quadrupole moment. Our results show that when including quadrupole moments in the training, the ability of the corresponding PiNet2-based ML models to infer MEP is improved significantly. This observation was further confirmed for the SPICE dataset. By regressing molecular multipoles, which are experimental observables, our work provides a generic strategy to infer MEP fast from ML models without explicit training on either MEP or electron density. 

\section{Methods}

\subsection{ML dipole and quadrupole models} 

In the atomic charge (AC) dipole model~\cite{Shao:2020kra, knijff_machine_2021}, atomic charges $q_i$, as latent variables, were predicted from pairwise interactions in a local environment. Then, they were used to fit the total dipole moment $\boldsymbol{\mu}$  calculated with Equation \ref{mu}:

\[\boldsymbol{\mu} = \sum_{i} q_i\cdot \mathbf{R}_i \label{mu} \tag{2},\]

where $\mathbf{R}_i$ is the position vector of atom $i$, and the summation runs over all atoms in the molecule. 

Alternatively, the atomic dipole (AD) model constructs the total dipole moment from  atomic dipole predictions  $\boldsymbol{\mu}_i$ by taking advantage of the equivariant neural network~\cite{pinet2}, as follows:

\[\boldsymbol{\mu} = \sum_{i} \boldsymbol{\mu}_i \label{ad_mu} \tag{3} \]

In this work, we introduce the AC quadrupole model, where the traceless quadrupole moment is calculated from atomic charges and coordinates using Equation \ref{Q}:

\begin{equation}
 \boldsymbol{Q}'_{\alpha\beta} = \sum_{i}{ q_i\cdot(3 R_{\alpha i}\cdot R_{\beta i } - \delta_{\alpha\beta}\cdot |\mathbf{R}_i|^2)},
 \label{Q}  \tag{4}
\end{equation}

where $\alpha$, $\beta$ $∈$ \{$x$, $y$, $z$\}.  

Given these base models, we further developed two dipole-quadrupole models:
\begin{itemize}
    \item AC-DQ model, where dipole and quadrupole moments were both derived from the same set of atomic charges by combing Equations \ref{mu} and \ref{Q} in the loss function with equal weights.
    \item AC-AD-DQ model, by combining Equations \ref{mu}, \ref{ad_mu}, \ref{Q} in the loss function with equal weights.
\end{itemize}

Finally, we also introduced two variants called AC-DQ-dw100 and AC-AD-DQ-dw100, with 100:1 being the dipole-quadrupole loss weight ratio. This comes with the consideration that quadrupole errors are usually much larger and would otherwise dominate training. All seven models used in this work are listed in Table~\ref{models}. In all these models, the net charge of each molecule was constrained to be zero. It is worth noting that the definition of the traceless quadrupole is not unique in the literature and one needs to be careful about the consistency between the model and the data. See the Appendix A for clarification. 

\begin{table}[H]
\centering
\caption{Different PiNet2-based ML models for predicting dipole and quadrupole investigated in this study. The weight ratio refers to the coefficients used for dipole and quadrupole terms in the loss function. AC: atomic charge; AD: atomic dipole; DQ: dipole-quadrupole; dw: dipole weight. }
\begin{tabular}{c c c l}
\toprule
Model & Dipoles& Quadrupoles& \makecell{Dipole-quadrupole \\ weight ratio} \\
\midrule
AC dipole& AC& --- & 1 : 0\\
AC-AD dipole& AC + AD & --- & 1 : 0\\
AC quadrupole& --- & AC & 0 : 1\\
AC-DQ& AC& AC&1 : 1\\
AC-DQ-dw100& AC& AC&1 : 100\\
AC-AD-DQ& AC + AD& AC&1 : 1\\
AC-AD-DQ-dw100& AC + AD& AC&1 : 100\\
\bottomrule
\end{tabular}
\label{models}
\end{table}

\subsection{The QM9 dataset and model training}

The QM9 dataset \cite{ramakrishnan_quantum_2014} consists of 133,885 stable organic molecules containing C, H, O, N and F atoms. To obtain the vector labels for molecular dipole moment and the labels for molecular quadrupole, we have computed these quantities for the QM9 dataset at the B3LYP/6-31G(2df,p) level of theory \cite{becke_density-functional_1993,6-31G}, using the NoSymmetry keyword in Gaussian~\cite{g09}. We used the vector or tensor rather than scalar multipole data and excluded 9,723 molecules that failed a consistency~\cite{Unke:2019bpa}. The Gaussian quadrupole moments were traceless and were multiplied by 3 for consistency with Equation \ref{Q} (see also the Appendix about different conventions of traceless quadrupole moments).  Note that the information block at the end of the Gaussian output file uses a third coordinate system when the NoSymmetry keyword is not used, which produces quadrupoles that are not always identical to the ones calculated from the input orientation. In addition, we computed the ESP charges for the QM9 dataset with Gaussian using the Merz-Kollman (MK) method~\cite{besler_merz_semiempirical, singh_kollman_esp}.  

Our previous work \cite{pinet2} showed that a large PiNet2 model was best for dipole predictions, therefore, the same model size was adopted in this work. All model-related hyper-parameters are listed in Appendix B. The dataset was split into training (80\%) and validation (20\%) sets, and the models were trained with 3M steps.  


\subsection{The SPICE dataset and model training}

The SPICE 2.0 dataset~\cite{eastman_spice_2023, eastman_nutmeg_2024}  was created to capture the interaction energies of proteins and small drug-like molecules. It includes molecular dynamics conformations of $\sim$114,000 structures composed of 17 elements (H, Li, B, C, N, O, F, Na, Mg, Si, P, S, Cl, K, Ca, Br, I) and has a good coverage of both the chemical and conformational space. We used SPICE for its inclusion of higher molecular weight and floppy molecules, which provide a more extended and complementary chemical space to the small organic molecule datasets, such as QM9. In SPICE, the energies and densities had been calculated using density functional theory (DFT), with the $\omega$B97M-D3(BJ) \cite{wB97M-V1, wB97M-V2} functional and the def2-TZVPPD \cite{TZVPPD1, TZVPPD2}  basis set \cite{eastman_spice_2023}. For our purpose, we have filtered the dataset to only include the lowest energy conformers of neutral, organic structures. The resulting subset contains 91,420 structures spanning 11 elements (H, B, C, N, O, F, P, S, Cl, Br and I).

Our initial tests showed that a medium size model is sufficient for training on the SPICE dataset, which was used subsequently. All model-related hyper-parameters are listed in the Appendix B. The dataset was split into training (80\%) and validation (20\%) sets, and the models were trained with 1M gradient descent steps. 

The chemical diversity of both the QM9 and SPICE datasets was visualised with 2048-bit Morgan fingerprints (radius=2) generated with RDKit~\cite{rdkit}. The molecular fingerprints were mapped with Uniform Manifold Approximation and Projection (UMAP) \cite{umapMcInnes2018}, using the Jaccard distance metric, a minimum distance 0.1 and 40 neighbours.  

\subsection{Assessing MEP from ML models}

The MEP from atomic charges was calculated with Equation \ref{Vi}: 
\[V_i(\mathbf{r}_i) =  \frac{1}{4\pi \epsilon_0}\sum_{j} \dfrac{q_j}{|\mathbf{r}_i - \mathbf{R}_j|} \label{Vi} \tag{5} \]
where $V_i$ is ESP at location $\mathbf{r}_i$, $q_j$ is the charge on nucleus $j$ and $|\mathbf{r}_i - \mathbf{R}_j|$ is the distance between evaluation point $i$ and nucleus $j$~\cite{bayly_well-behaved_1993}. This provides a consistent way to compare the ESP charges and atomic charges predicted from ML models on producing MEPs for the QM9 dataset.

In this case, the molecular vdW surfaces for the validation set were constructed by placing a sphere with a Bondi vdW radius~\cite{Bondi_VdW} around each atom. These atomic spheres were combined to form the molecular vdW surface and uniformly covered with points using the Fibonacci lattice method. About 4400 points were included for each molecule, which corresponds to a point density of 114~\AA$^{-2}$. Then, the electrostatic potential at each surface point was evaluated using Equation \ref{Vi}.

As a representative example of the SPICE dataset, the structure of a tryptophan derivative N-acetyl-L-tryptophan methylamide was taken with adjacent water molecules removed. The MEP of this molecule was calculated at the same DFT level of theory used in the original SPICE dataset and on the 0.001 a.u. electron density isosurface with Psi4~\cite{psi4} and Multiwfn~\cite{lu_multiwfn_2012, lu_comprehensive_2024, zhang_efficient_2021}. The same isosurface was then used in Equation \ref{Vi} for obtaining the MEP from the atomic charges, which allows a fair comparison to that obtained from the DFT charge density. A similar procedure was applied for the showcase of MEPs of both propylene carbonate and fluoropropylene carbonate molecules. 

\section{Results}

\subsection{Performance of ML models for dipole and quadrupole predictions on QM9}

 The accuracies in predicting dipoles and quadrupoles with the ML models are shown in Table~\ref{tab:loss}. The performance of AC dipole model is reproducible as compared to the previous work~\cite{pinet2} and the same applies to the AC-AD dipole model. For the new AC quadrupole model, we obtained a mean absolute error (MAE) per tensor component of 0.147 D$\cdot$\AA. This can be compared to the reported MAE of 0.221 D$\cdot$\AA~for a very similar dataset in a recent work of XPaiNN with higher-order geometrical features~\cite{10.1021/jacs.5c12428}.  To put these numbers into perspective, the MAE per tensor component between coupled-cluster singles and doubles (CCSD) calculations and experiments is about 0.242 D$\cdot$\AA~\cite{10.1063/1.4951685} (Note that the original publication used the Buckingham convention for the traceless quadrupole and we did the factor conversion for a fair comparison, see Appendix A for clarification) . Therefore, the PiNet2-based AC quadrupole model shows an excellent performance that reaches the chemical accuracy.

 When combining the AC dipole and AC quadrupole into the AC-DQ model, the dipole predictions worsen slightly, and the quadrupole predictions are practically unchanged. Because the loss of the AC quadrupole model is two orders of magnitude higher than that of the AC dipole model, the quadrupole losses dominated training if the weights were not adjusted accordingly. Therefore, we increased the weight ratio between dipole-quadrupole loss by a factor of 100, denoted as dw100. The AC-DQ-dw100 with the increased dipole weight leads to more balanced predictions as one would expect. Surprisingly, the quadrupole MAE of the AC-DQ-dw100 model becomes even smaller than that of the AC quadrupole model. Moreover, it is worth noting that the validation curve follows the training curve more closely when the dipole contribution is included in the training data. 
 
 Including the atomic dipole contribution gives the models more flexibility because both multipoles are not constrained by the same set of atomic charges. Nevertheless, this change did not improve the predictions in practice. The AC-AD-DQ models performed worse than the AC-DQ models and were more prone to overfitting. Adding a dipole contribution in the dw100 models slightly increased model accuracy. Overall, the AC-DQ-dw100 model shows the best performance among the dipole-quadruple models. 
 
\begin{table}[H]
    \caption{Dipole and quadrupole mean absolute errors (MAEs) of the different ML models based on PiNet2 architecture in the validation set of the QM9. The errors are given per tensor component.}
\begin{center}
\begin{tabular}{c  c c c c c}
\toprule
Model &
\makecell{Dipole\\MAE / D} &
\makecell{Rel. dipole\\MAE} &
\makecell{Quad.\\MAE / D$\cdot$\AA} &
\makecell{Rel. quad.\\MAE} \\
\midrule
 AC dipole& 0.013&1.0 & --- &---\\
 AC-AD dipole& 0.013&1.0 & ---&---\\
 AC quadrupole& ---&---& 0.147
&1.0\\  
AC-DQ& 0.028& 2.2 & 0.144
&1.0
\\
AC-DQ-dw100& 0.018& 1.4 & 0.137
& 0.9\\
AC-AD-DQ& 0.032& 2.5& 0.171
& 1.2
\\
 AC-AD-DQ-dw100 &0.019 &1.4 & 0.150& 1.0\\
\bottomrule
\end{tabular}
    \label{tab:loss}
\end{center}
\end{table}

\subsection{Inference of atomic charges on QM9}

Before looking into the results regarding the MEP produced by atomic charges from various ML models, it would be interesting to begin by examining their element-wise distributions, which provide a first impression about how similar these charges are.

\begin{figure}[H]
 \centering
        \includegraphics[width=1.0\textwidth]{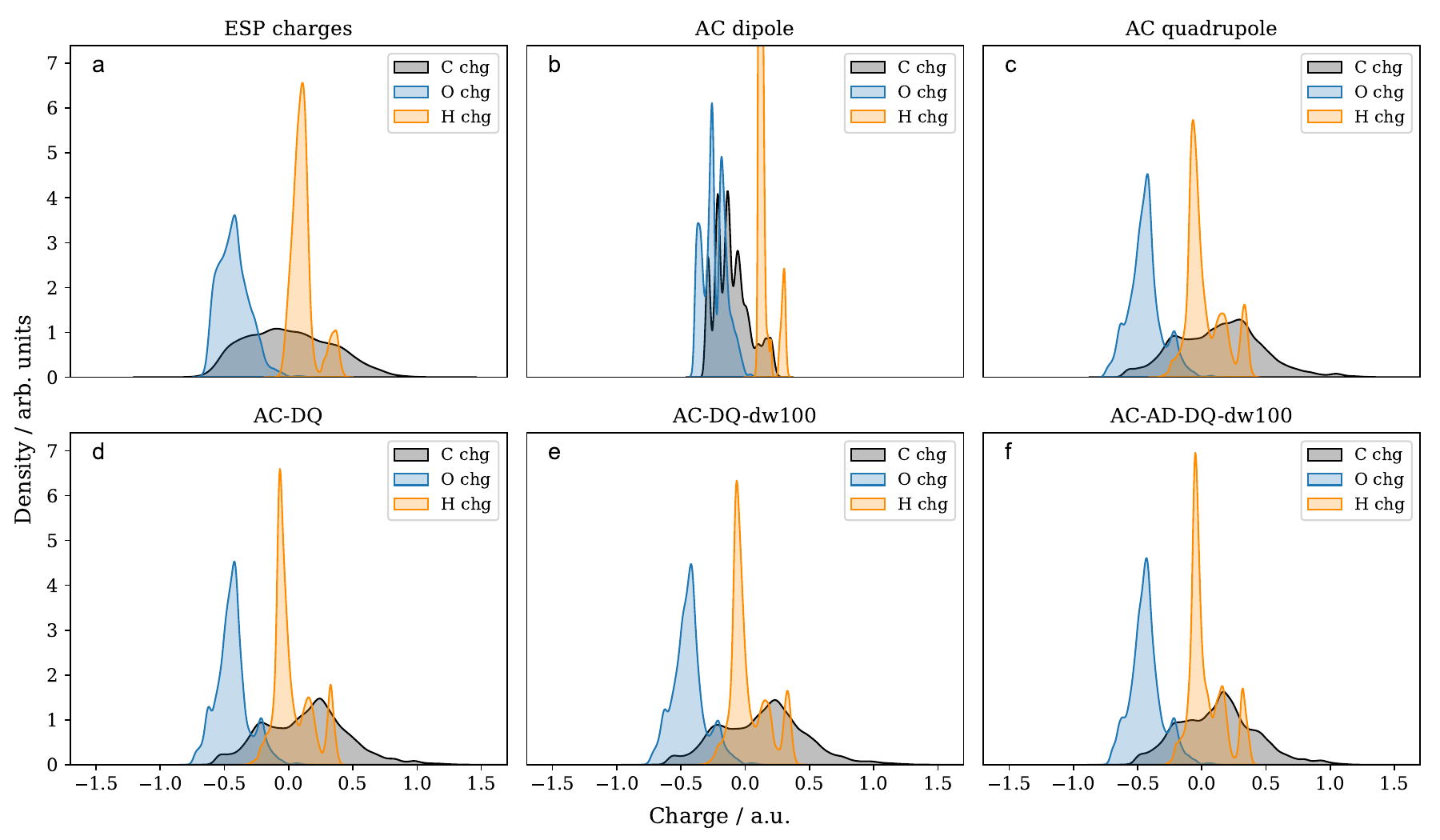}
        \caption{Density distributions of atomic charges predicted by different PiNet2-based ML models on the QM9 validation set.}
\label{fig:charges}
\end{figure}

Compared to the ESP charges, which are obtained from fitting the MEP directly, the AC dipole model produces a much narrower charge distribution with charges concentrated into sharp peaks (Figure \ref{fig:charges}). Instead, the AC quadrupole and other dipole-quadrupole models produce more dispersed distributions, which highly resemble that of ESP charges. It is striking to see that the quadrupole moment dictates the overall shape of the charge distribution regardless of the weight ratio between the dipole and quadrupole contributions.  

On the other hand, it is found that the AC dipole charges actually show a smaller MAE (also RMSE) with respect to the ESP charge as compared to AC quadrupole and dipole-quadrupole models (see Table~\ref{tab:mae-rmse} in the Appendix C). This is because the charge distribution from the AC dipole model is concentrated around zero, which makes the error metrics looks better. Therefore, the typical error metrics in this case failed to evaluate the physical similarity (performance) between models, which was also reported in the development of machine learning interatomic potentials (MLIPs)~\cite{2021.Kovacs}. 

\subsection{Inference of electrostatic potentials on QM9}

We then evaluated the MEP generated from different atomic charges.  Here, we compared the performance of atomic charges predicted by different models on the vdW surface around the molecules in the QM9 validation set, as described in the methods section. The reference potential was obtained from ESP charges, and the potential from atomic charges was calculated with Equation~\ref{Vi}.

Comparing AC dipole and AC quadrupole models, the MEP from the AC quadrupole model is significantly better in reproducing the MEP from the ESP charges (see Table \ref{tab:test_set_esp_performance}). The dipole-quadrupole models enjoy similar improvement in reproducing MEPs once the quadrupole contribution is introduced into the models. These observations agree very well with the general trend observed in the charge distribution shown and discussed in the previous section. Interestingly, the AC-AD-DQ types of models, which performed worse on dipole and quadrupole predictions, appear to champion the chart on reproducing MEPs with a small margin.

\begin{table}[H]
\centering
\caption{Performance of the atomic charges from PiNet2 models in predicting ESP from ESP charges on the approximate van der Waals surface of the molecules in the QM9 validation set. The reference was the potential generated by ESP charges.}
\small
\begin{tabular}{c c c }
\toprule
 &Pearson R&RMSE / a.u.\\ 
 \midrule
 AC dipole &0.916&0.012\\
 AC-AD dipole& 0.629&0.026\\
 AC quadrupole& 0.954&0.0090\\
 AC-DQ &0.955& 0.0089\\ 
 AC-DQ-dw100&0.954& 0.0090\\ 
 AC-AD-DQ& 0.960& 0.0083\\
 AC-AD-DQ-dw100& 0.961&0.0081\\
 \bottomrule
\end{tabular}
\label{tab:test_set_esp_performance}
\end{table}

\begin{figure}[H]
    \centering
    \includegraphics[width=0.8\textwidth]{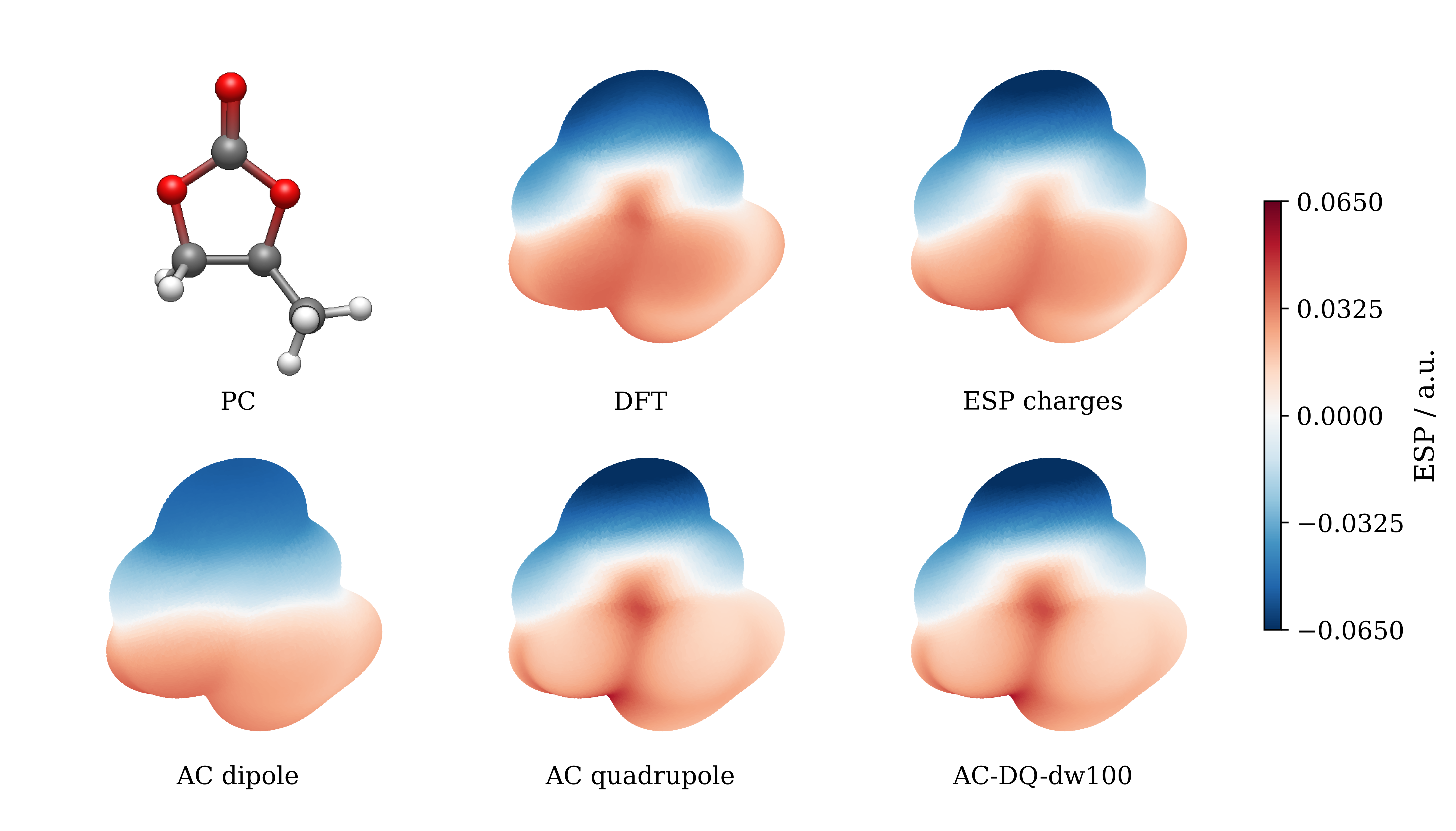}
\caption{Electrostatic potential of propylene carbonate calculated on the 0.001 a.u. density isosurface. The bottom panel shows the potential from atomic charges predicted with PiNet2-based dipole and quadrupole models.}
\label{fig:python_esp1}
\end{figure}

\begin{figure}[H]
    \centering
    \includegraphics[width=0.8\textwidth]{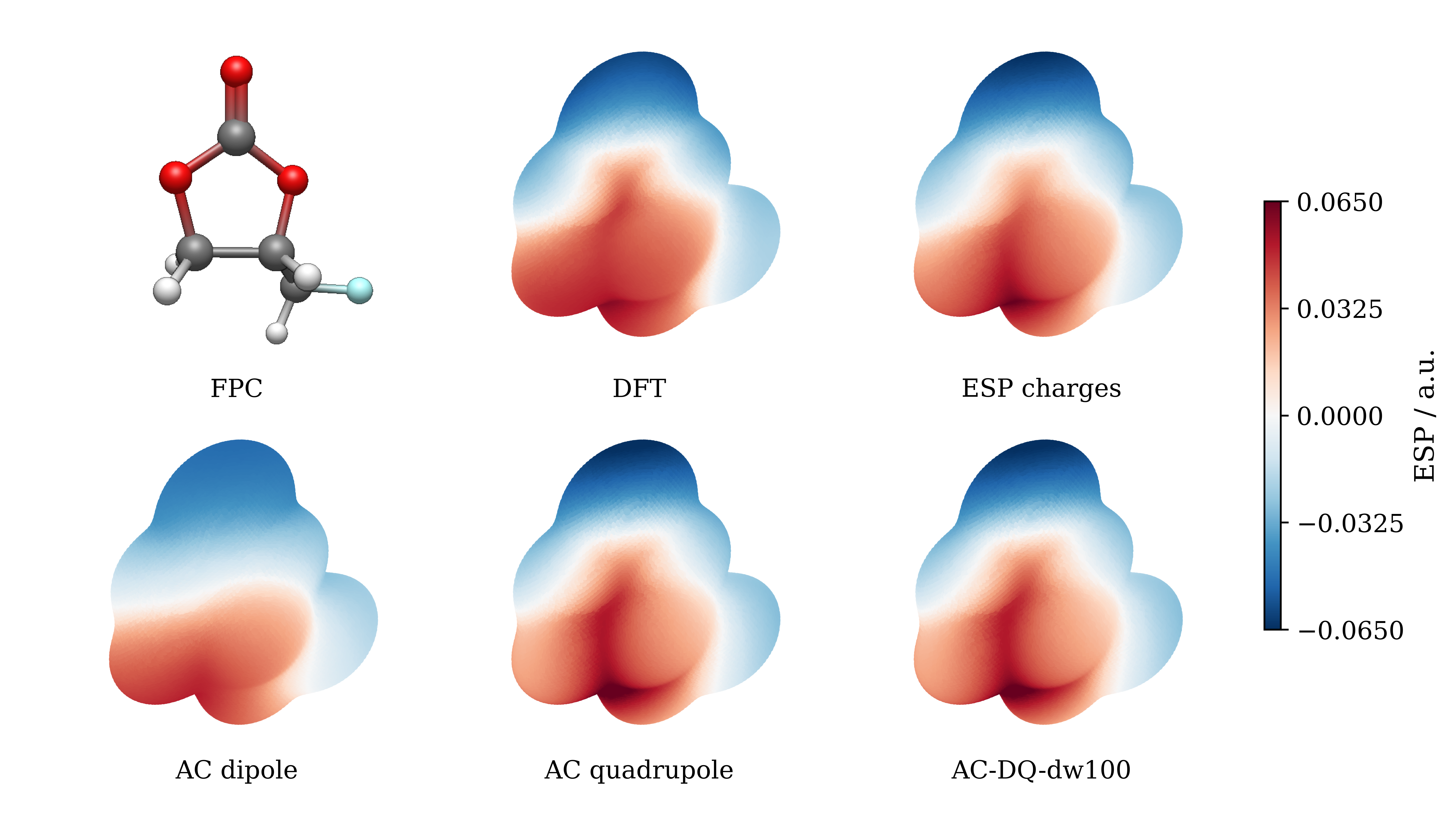}
\caption{Electrostatic potential of fluoropropylene carbonate calculated on the 0.001 a.u. density isosurface. The bottom panel shows the potential from atomic charges predicted with PiNet2-based dipole and quadrupole models.}
\label{fig:python_esp2}
\end{figure}

To visualize these differences in reproducing MEPs from different ML models, we have chosen two popular solvent molecules in electrochemical energy storage: propylene carbonate and fluoropropylene carbonate. As shown in Figures~\ref{fig:python_esp1} and~\ref{fig:python_esp2}, the charges predicted by the AC quadrupole and dipole-quadrupole models can quantitatively reproduce MEP on the van der Waals surface of these molecules. Instead, the MEPs inferred from the AC dipole model provide only a qualitative picture. 

\subsection{Extension to the SPICE dataset}

Based on the investigations on QM9 shown above, we have chosen three models AC dipole, AC quadrupole and AC-DQ-dw100 to see their performances on the SPICE dataset, as they represents the two extremes and their best mixture (Table~\ref{tab:loss}). 

As shown in Table~\ref{tab:spice_loss}, the AC-DQ-dw100 lowers the prediction error for the quadrupole moment and shows a very similar performance on predicting dipole moment as compared to the AC dipole moment. This is exactly the same as we have observed for the QM9 dataset (Table~\ref{tab:loss}). Compared to the prediction errors for the QM9 dataset, the corresponding ones for the SPICE dataset are about one order of magnitude larger. This may look striking at first glance; however, similar prediction accuracy has been reported with other types of equivariant graph-convolutional neural network architecture, such as MACE trained on the SPICE dataset~\cite{2025.Kovacs}. 

\begin{table}[H]
\begin{center}
\caption{Dipole and quadrupole mean absolute errors (MAE) of the different ML models based on
PiNet2 architecture trained on the filtered SPICE dataset.}
\begin{tabular}{c  c c c c c}
\toprule
Model &
\makecell{Dipole\\MAE / D} &
\makecell{Rel. dipole\\MAE} &
\makecell{Quad.  \\MAE / D·Å} &
\makecell{Rel. quad.\\MAE} \\
\midrule
 AC dipole& 0.28 &1.0 & ---&---\\
AC quadrupole& ---&---& 2.28  &1.0\\
AC-DQ-dw100& 0.30  & 1.0 & 2.21 & 1.0\\
\bottomrule
\end{tabular}
\label{tab:spice_loss}
\end{center}
\end{table}

\begin{figure}[H]
    \centering
    \includegraphics[width=1.0\textwidth]{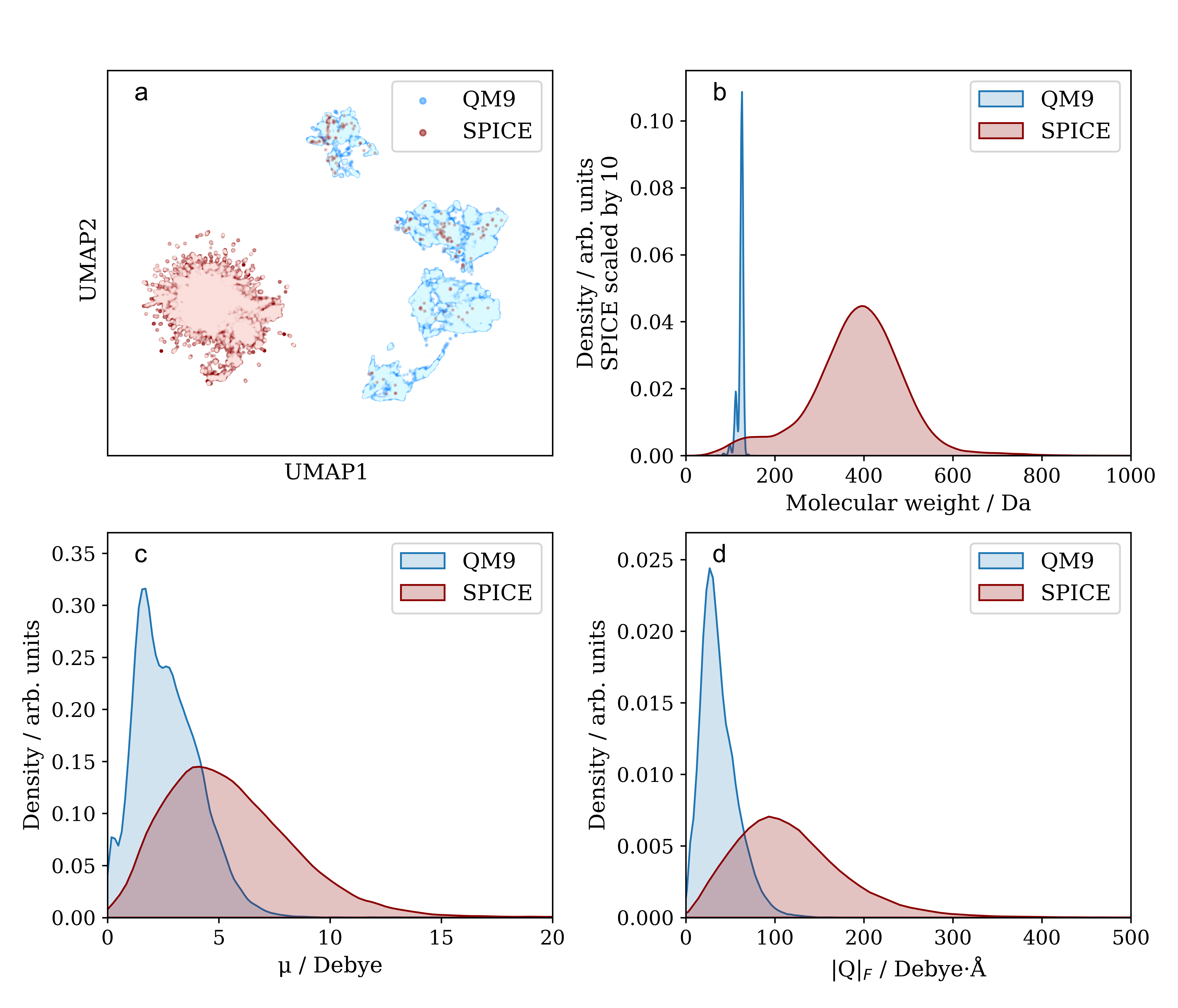}
    \caption{a) Embedding of the QM9 and SPICE datasets using Morgan structural fingerprints. The validation sets are lighter in colour and were mapped on top. Molecules were mapped according to similarity, with more similar structures clustered together; b) The comparison of the molecular weights (MWs) between two datasets; c) The comparison of the total dipole distributions between these two datasets; c) The comparison of the total quadrupole distributions between these two datasets.}
    \label{fig:umap}
\end{figure}

To better understand the intrinsic differences between the QM9 and the SPICE datasets, we have analysed their respective chemical space. As shown in Figure~\ref{fig:umap}a, the SPICE and QM9 datasets occupy different chemical spaces. In addition, SPICE has a significantly wider distribution of molecular weights, as shown as compared to QM9 (Figure~\ref{fig:umap}a). This arises in part from the inclusion of larger molecules and heavier elements in the SPICE dataset, as well as from the presence of dimers or solvating water molecules in some structures. On the same note, the filtered SPICE dataset included 11 chemical elements compared to 5 for QM9.

These difference found in the projected chemical space and in the distribution of molecular weights also manifest themselves in the molecular dipole and the quadrupole moment distributions. As shown in Figures~\ref{fig:umap}c and~\ref{fig:umap}d, the dipole and quadrupole distributions are correspondingly wider for SPICE as compared to those for QM9. We consider these intrinsic differences between the datasets as the main cause for the very different prediction errors presented in Table~\ref{tab:loss} and Table~\ref{tab:spice_loss}. 

Finally, we have also checked the MEP in the SPICE dataset. Given the size of the SPICE dataset and the molecular weights of molecules therein, we were prohibited to compute the reference MEP for all molecules in the validation set, as we did and showed for QM9 in Table~\ref{tab:test_set_esp_performance}. Instead, we have chosen N-acetyl-L-tryptophan methylamide as a representative for showcase. For large and floppy molecules, the ability of the ESP charges to reproduce the MEP can be questioned. Therefore, the reference MEP shown in Table~\ref{tab:esp_performance_spice} is the MEP from DFT electron density instead. Nevertheless, it is found that the main observation that we made for the QM9 dataset remains true for this case: the inclusion of the quadrupole contribution into the ML models significantly improves their ability to infer MEP as compared to the AC dipole model.

\begin{table}[H]
\small
\centering
\caption{Performance of the atomic charges from dipole-quadrupole models trained on SPICE in predicting ESP on the 0.001 a.u. density isosurface of N-acetyl-L-tryptophan methylamide}
\begin{tabular}{c c c c c}
\toprule
& Pearsonr& RMSE  / a.u. & min ESP / a.u. & max ESP / a.u. \\
\midrule
DFT &   1 &  0  & -0.0755& 0.0826\\
ESP charges &  0.995& 0.00470& -0.0890& 0.0753\\
AC dipole&  0.967& 0.00960& -0.0716& 0.0794\\
AC quadrupole& 0.986& 0.00634& -0.0856& 0.0763\\
AC-DQ-dw100&  0.983& 0.00705& -0.0879& 0.0816\\
\bottomrule
\end{tabular}
\label{tab:esp_performance_spice}
\end{table}

\section{Conclusion and outlook}

The standard view from the multipole expansion (Equation~\ref{multipole_exp}) is that, for neutral molecules, the dipole moment is the leading term. Consistent with this, ESP charges, which are designed to reproduce the MEP, yield dipole moments in excellent agreement with reference DFT calculations (Appendix D).

However, in this work we show that atomic charges chosen to reproduce the quadrupole moment in ML models perform better at recovering the MEP than charges fitted to reproduce the dipole moment. When both dipole and quadrupole contributions are included in the models, the quality of the quadrupole fit is further improved, albeit at the expense of the dipole moment accuracy. These two observations hold for both the QM9 and SPICE datasets, which sample complementary regions of the chemical space of organic molecules.

Because there is no one-to-one mapping between the MEP and an atomic charge partitioning, our results suggest that the quadrupole moment is a more suitable primary fitting target for ML-based atomic charge models than the dipole moment alone. The quadrupole moment shares the advantageous status of being an experimental observable, yet it is far more compact than the full charge density in terms of the information that must be stored or predicted. Taken together, our findings provide a practical strategy for obtaining rapid, ML-based access to the MEP, which is crucial for data-driven solvent and electrolyte design~\cite{10.1002/batt.202000262}.

\ack{This project has received funding from the European Research Council (ERC) under the European Union's Horizon 2020 research and innovation programme (grant agreement No. 949012). This work was partially supported by the Wallenberg Initiative Materials Science for Sustainability (WISE) funded by the Knut and Alice Wallenberg Foundation (KAW) and the Swedish Energy Agency
(Project no. P2025-03176). The calculations were performed on the resources provided by the National Academic Infrastructure for Supercomputing in Sweden (NAISS) at C3SE and UPPMAX, partially funded by the Swedish Research Council through grant agreement No. 2022-06725.}

\roles{Kadri Muuga: Data curation; Formal analysis; Investigation; Methodology; Project administration; Writing - original draft; Writing - review \& editing; Lisanne Knijff: Data curation; Formal analysis; Investigation; Methodology; Writing - review \& editing; Chao Zhang: Conceptualization; Funding acquisition; Investigation; Methodology; Project administration; Resources; Supervision; Writing - original draft; Writing - review \& editing.}

\data{The data that support the findings of this study are available from the corresponding authors upon reasonable request. The QM9 quadrupole data is available on the Figshare: link made available upon publication.}

\appendix

\section{Traceless quadrupole definitions}

The first definition, or the Buckingham convention, is given by the IUPAC~\cite{renner2007quantities}

\begin{equation}
    Q'_{ij} = \frac{1}{2}(3Q_{ij} - \mathrm{tr}(Q)\delta_{ij})
\end{equation}

This definition is commonly used when comparing to an experimental reference. 

The second definition is expressed as

\begin{equation}
    Q'_{ij} = 3Q_{ij} - \mathrm{tr}(Q)\delta_{ij}
\end{equation}

This definition is listed by Jackson's \emph{Classical Electrodynamics}~\cite{jackson_classical_1999}, and adopted in this work.

Finally, the third definition can be written as

\begin{equation}
    Q'_{ij} = Q_{ij} - \mathrm{tr}(Q)\delta_{ij}/3
\end{equation}

This is the definition used by the Gaussian~\cite{g09} software.

\section{Network parameters}

\begin{table}[H]
\begin{center}
    \caption{Hyperparameters used to train the PiNet2-dipole and dipole-quadrupole models for the QM9 dataset. $R_\textrm{c}$ stands for the interaction cut-off, graph convolution (GC) blocks is the model depth and $n_\textrm{basis}$ is the number of basis functions.}
\begin{tabular}{c c c c }
\toprule
  Layer&Architecture&Parameter&Value\\ 
 \midrule
  PI&[64] × 10&$R_\textrm{c}$&4.5 \AA\\  
 II&[64, 64, 64, 64]&GC blocks&5\\
 IP& [64, 64, 64, 64]& $n_\textrm{basis}$&10\\
 Output&[64, 1]&&\\
 \bottomrule
 \end{tabular}
    \label{tab:hyperparameters}
\end{center}
\end{table}

\begin{table}[H]
\begin{center}
    \caption{Hyperparameters used to train the PiNet2-dipole and dipole-quadrupole models for the SPICE dataset.  $R_\textrm{c}$ stands for the interaction cut-off, GC blocks is the model depth and $n_\textrm{basis}$ is the number of basis functions.}
\begin{tabular}{c c c c }
\toprule
  Layer&Architecture&Parameter&Value\\ 
 \midrule
  PI&[32] × 10&$R_\textrm{c}$&4.5 \AA\\  
 II&[32, 32, 32, 32]&GC blocks&5\\
 IP& [32, 32, 32, 32]& $n_\textrm{basis}$&10\\
 Output&[32, 1]&&\\
 \bottomrule
 \end{tabular}
    \label{tab:hyperparameters2}
\end{center}
\end{table}

The dataset was split into training (80\%) and validation sets (20\%). A batch size of 100 samples was used for the training. The Adam optimizer~\cite{kingma2017adam} was used with a learning rate of 0.0001, which decayed by a factor of 0.994 every 10,000 steps. A global gradient norm clipping strategy was adopted to prevent exploding gradients. Training was terminated after 3 million  gradient descent steps for the QM9 dataset and 1 million gradient descent steps for SPICE.

\section{Error metrics for the charge distribution comparisons}

\begin{table}[H]
\centering
\caption{Mean Absolute Error (MAE) and Root Mean Square Error (RMSE) of predicted charges relative to the ESP charges from different PiNet2 -based ML models on the QM9 validation set.}
\begin{tabular}{c c c}
\toprule
&MAE /\textit{e}&RMSE / \textit{e}\\
\midrule
AC dipole&0.129&0.184\\
AC-AD dipole & 0.217& 0.282\\
AC quadrupole & 0.170& 0.254\\
AC-DQ&0.168&0.250\\
AC-DQ-dw100&0.170&0.252\\
AC-AD-DQ& 0.158&0.237\\
AC-AD-DQ-dw100&0.154&0.230\\
\bottomrule
\end{tabular}
\label{tab:mae-rmse}
\end{table}

\section{ESP charges for reproducing molecular dipole and quadrupole}

\begin{figure}[H]
    \centering
    \includegraphics[width=0.9\linewidth]{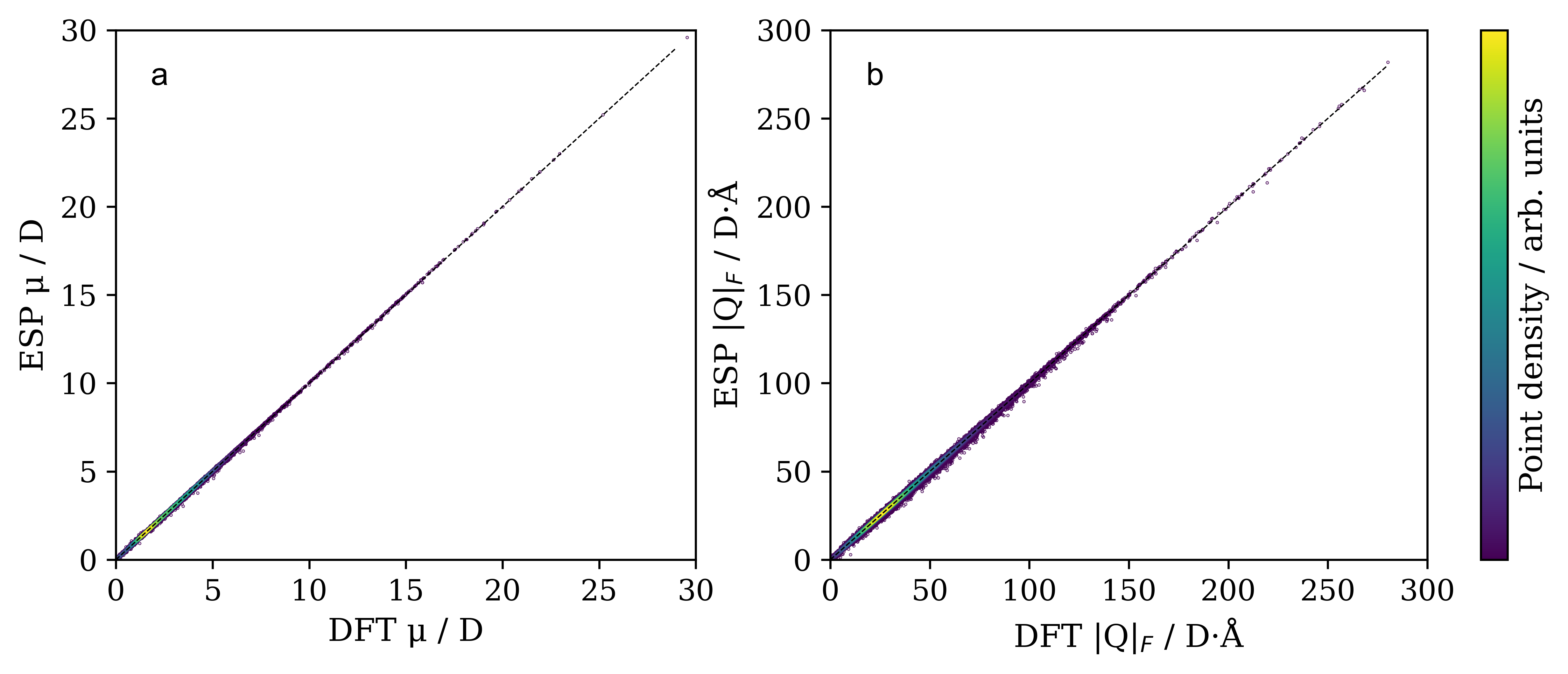}
    \caption{The correlation of (a) dipole and (b) quadrupole magnitudes obtained from MK ESP charges with DFT reference values for the QM9 dataset. The dipole and quadrupole moments were calculated from ESP charges using Equations \ref{mu} and \ref{Q}, respectively. All properties are reported at the same level of theory as used in the original QM9 dataset. The per-tensor-component dipole MAE from ESP charges is 0.020 D and that for quadrupole is 0.323 D$\cdot$Å. }
    \label{fig:D_Q_from_esp}
    \end{figure}

\bibliographystyle{ieeetr}

\end{document}